# Astro2020 Science White Paper

# Empirically Constraining Galaxy Evolution

**Thematic Areas:** Galaxy Evolution

**Principal Author:**
Name: Peter Behroozi
Institution: University of Arizona
Email: behroozi@email.arizona.edu
Phone: (520) 621-9629

**Co-authors:**
**Matthew R. Becker** (Argonne), **Frank C. van den Bosch** (Yale), **Jarle Brinchmann** (Leiden), **Charlie Conroy** (Harvard University), **Mark Dickinson** (NOAO), **Christopher M. Hirata** (OSU), **Andrew Hearin** (Argonne), **Alexie Leauthaud** (UCSC), **Chun Ly** (University of Arizona), **Yao-Yuan Mao** (University of Pittsburgh), **Benjamin P. Moster** (LMU), **Christine O'Donnell** (University of Arizona), **Casey Papovich** (Texas A&M University), **Aldo Rodríguez-Puebla** (UNAM), **Rachel Somerville** (CCA, Flatiron Institute), **Erik Tollerud** (STScI), **Jeremy Tinker** (NYU), **Yun Wang** (Caltech/IPAC), **Risa H. Wechsler** (KIPAC; Stanford; SLAC), **Charity Woodrum** (University of Arizona), **Ann Zabludoff** (University of Arizona), **Dennis Zaritsky** (University of Arizona), **Andrew R. Zentner** (University of Pittsburgh), **Huanian Zhang** (University of Arizona)

**Abstract**:
Over the past decade, empirical constraints on the galaxy—dark matter halo connection have significantly advanced our understanding of galaxy evolution. Past techniques have focused on connections between halo properties and galaxy stellar mass and/or star formation rates. Empirical techniques in the next decade will link halo assembly histories with galaxies' circumgalactic media, supermassive black holes, morphologies, kinematics, sizes, colors, metallicities, and transient rates. Uncovering these links will resolve many critical uncertainties in galaxy formation and will enable much higher-fidelity mock catalogs essential for interpreting observations. Achieving these results will require broader and deeper spectroscopic coverage of galaxies and their circumgalactic media; survey teams will also need to meet several criteria (cross-comparisons, public access, and covariance matrices) to facilitate combining data across different surveys. Acting on these recommendations will continue enabling dramatic progress in both empirical modeling and galaxy evolution for the next decade.

**CONTEXT**

In the ΛCDM paradigm, galaxies form at the centers of halos, which are self-bound, virialized regions of dark matter. Understanding galaxy formation has been a central theme of astronomy for over a century. This is due both to the rich astrophysics of radiative magnetohydrodynamics in galaxy formation as well as galaxies' relevance for many other fields. For example, most modern probes of dark energy use galaxies as tracers to infer the underlying dark matter distribution (e.g., weak lensing, baryonic acoustic oscillations, cluster cosmology), and precision cosmology requires understanding biases and uncertainties in the galaxy—matter link. Galaxies' formation histories also affect many smaller-scale physical processes—e.g., supernovae and gamma-ray bursts, supermassive black hole evolution and merging, dust production and destruction, dynamics of stars and the interstellar medium, and even planet formation histories. **Hence, a better understanding of galaxy formation carries benefits all across astronomy.**

We have made enormous progress in understanding galaxy formation in halos (see e.g., Wechsler & Tinker 2018). In the past decade, greater computing power has allowed higher-fidelity simulations, which in turn yielded more productive communication between observers and theorists. **At the same time, a new class of empirical models gave us the ability to transform observational snapshots of galaxies into movies of their evolution, as well as the ability to perform self-consistent joint inference from multiple classes of observations.** As in cosmology, where combining observations from multiple experiments has dramatically reduced uncertainties, empirical models have broken degeneracies in galaxy formation in a powerful and extremely cost-effective manner. Empirical models also allow theorists to accurately predict and marginalize over what observers would see, with the result that every large survey project or major telescope (e.g., *JWST*, DES, DESI, LSST, *WFIRST*, etc.) has empirical mock catalogs that have played (or will play) key roles in survey design, validation, and interpretation.

The most familiar empirical models involve the galaxy stellar mass—dark matter halo mass relation. These models often combine constraints on the observed number densities of galaxies with the simulated number densities of dark matter halos (see references for Fig. 1). Counting galaxies is much easier than directly measuring their halo masses, so this technique has allowed determining galaxies' dark matter halo masses across a much wider range of redshifts and stellar masses than possible with any other technique (see Wechsler & Tinker 2018, for a review). **The promise of empirical models goes far beyond these initial successes.** As evidence, we point to initial empirical models of galaxies' circumgalactic media (CGM; Zhang et al. 2018a, 2018b), cold gas disks (Popping et al. 2015), sizes (Rodríguez-Puebla et al. 2017, Somerville et al. 2018), supermassive black holes (Haowen Zhang et al., in prep), morphologies (Rodríguez-Puebla et al. 2017), colors (Behroozi et al. 2018), metallicities (Rodríguez-Puebla et al. 2016), short gamma-ray burst rates (Behroozi et al. 2014), and contribution to reionization (Finkelstein et al. 2019). Recent progress in models that connect individual halo formation histories to individual galaxy formation histories (e.g., Moster et al. 2018, Behroozi et al. 2018) has now made it possible to generate these galaxy properties in a much more general and assumption-free manner, providing a path for these techniques to become established research approaches in the next decade. The promise of empirical models and the support necessary to ensure their success are the main subjects of this white paper.

# INTRODUCTION TO EMPIRICAL MODELS

Empirical models have been used for centuries to accelerate our understanding of physical laws. For example, Kepler's Laws of orbital motion (i.e., empirical relationships between orbital properties) were a key inspiration for Newton's Law of Gravity. **Empirical galaxy formation models yield connections between galaxies, their surrounding gas, and their dark matter halos that help clarify the underlying responsible physics.**

As an example, the median relationship between galaxies' stellar masses and their host halo masses can be expressed as a function $f$, such that

median galaxy stellar mass = $f$ (host halo mass, redshift)

The true functional form of $f$ is not known, but any suitably flexible function (e.g., a double power law or spline) can approximate $f$. Hence, different parameterizations for $f$ nonetheless give very similar conclusions (Fig. 1). Any choice of $f$ specifies how galaxies should populate a dark matter simulation. Hence, modelers apply a trial choice of $f$ to a dark matter simulation to generate a mock Universe, which is then compared to real observations (e.g., to galaxy number densities and correlation functions) to find discrepancies. Via an MCMC algorithm, many trial choices are tested to converge on the closest form of $f$ for the real Universe (Fig. 2).

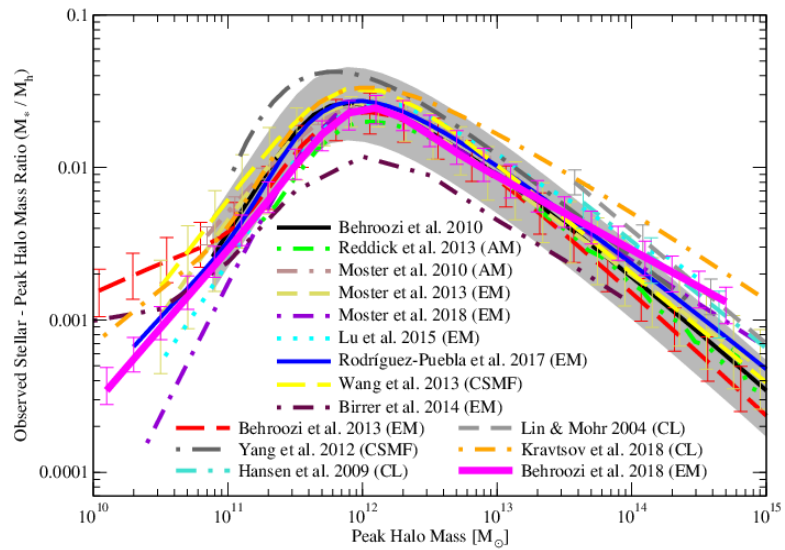

FIG 1. The galaxy stellar mass—halo mass relation at $z \sim 0$, as derived by many different techniques, from Behroozi et al. (2018). See reference list for other references cited in the figure. Errors are dominated by systematic ~0.2-0.3 dex uncertainties in galaxy stellar masses.

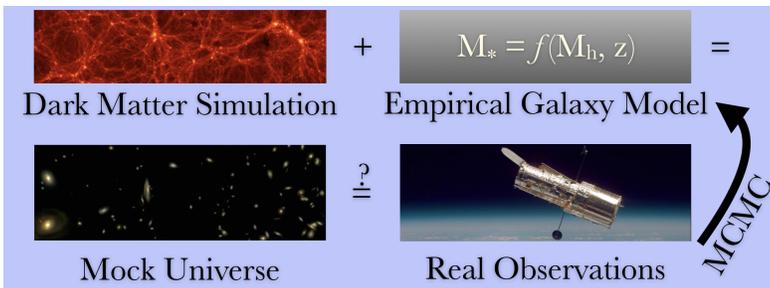

FIG 2. Basic schematic for empirical modeling. A dark matter simulation (top left) gives halo locations and properties. For any galaxy—halo relationship (top right), one can populate simulated halos with galaxies, generating a mock Universe (bottom left). This is compared with real observations (bottom right), and the galaxy—halo relationship is iteratively updated until galaxy statistics in the mock and real Universes converge. Images used are cited in the references section.

Importantly, $f$ derived in this way does not have any explicit dependence on gas physics, turbulence, etc. Instead, it is the emergent relation between stellar mass and halo mass that is the product of all these physical processes in the real Universe. The outcome is <u>independent</u> of our understanding (or lack thereof) of gas physics, meaning that *f* can reveal physics that was not expected *a priori*. For example, the lack of evolution in *f* from $z=4$ to $z=0$ has severely constrained the influence of factors other than halo mass for influencing how efficiently incoming baryons are converted to stars (Behroozi et al. 2013). In addition, hydrodynamical simulations attempting to match these constraints found a universal need for early, strong stellar feedback that had not previously been included (Hopkins et al. 2014, Wang et al. 2015).

**Empirical models have provided the unique ability to link galaxies across time to create movies of galaxy evolution (Fig. 3), which is critical to inform better physical models.** This ability arises from self-consistently combining different observational data sets. Different average galaxy evolutionary histories will result in different galaxy number densities as a function of redshift—e.g., faster evolution results in fewer galaxies at earlier redshifts. Since a given mock Universe can be observed at many different redshifts (with selection effects that mimic different surveys), observed data sets at many different redshifts can be combined to yield constraints on galaxy evolutionary histories. The general ability to break degeneracies (e.g., the relative impact of host halo mass vs. baryonic accretion on galaxy quenching) by combining multiple data sets (e.g., the evolution of galaxy quenched fractions and the spatial correlation of quenched vs. star-forming galaxies) is also widely used in cosmology, as no single observable is sufficient to constrain all cosmological parameters simultaneously.

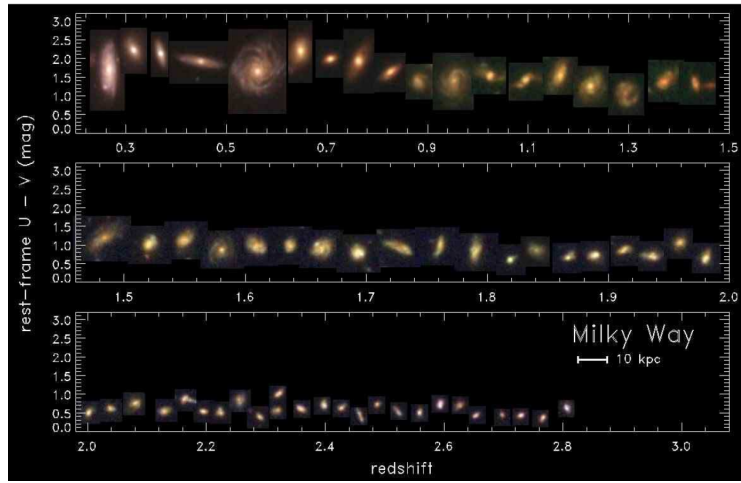

FIG 3. "Movie" of galaxy formation from Papovich et al. (2015).

## FUTURE SCIENCE

**Empirical models are not limited to the stellar mass—halo mass relation.** One may look for empirical relationships between *any* galaxy property and any combination of halo and other galaxy properties. To the extent that observational data can be combined to constrain this relationship, then an empirical model can help clarify the underlying physics. At the same time, this empirical model can be used to generate mock catalogs that are both consistent with all observed data used as constraints *and* include new information (such as host halo properties) that is not practically accessible by direct observation. A few examples:

**Galaxy star formation history** $= f_{SFH}$(halo growth history, environment, redshift)
**Galaxy size** $= f_{size}$(halo radius, galaxy star formation rate, environment, redshift)
**Galaxy morphology** $= f_{shape}$(star formation history, environment, halo merger history, redshift)
**Galaxy gas content** $= f_{gas}$(halo growth history, galaxy star formation rate, environment, redshift)
**Galaxy supermassive black hole mass** $= f_{BH}$(galaxy mass, galaxy morphology, redshift)
**Galaxy SNe/GRB/FRB rate** $= f_{transient}$(galaxy star formation & metallicity history)

Many more are possible (e.g., metallicity, temperature & clumpiness of the CGM, and galaxy colors). In each case, an empirical relationship can be written down between the desired galaxy property and properties that are known from the underlying dark matter simulation and/or accessible from another empirical model. Crucially, all such relationships allow translating the known evolution of dark matter halos into knowledge about the evolution of galaxy properties. The most critical questions in galaxy formation (e.g., why do galaxies stop forming stars?, why do galaxies change shape?, what is the relationship between supermassive black holes and galaxies?) *all* involve how galaxies change with time. **Resolving the most important galaxy formation questions becomes dramatically easier with empirical models' ability to convert static observed galaxies into movies of how galaxies evolve.**

**OBSERVATIONAL & DATA ANALYSIS NEEDS:**

Many planned surveys/telescopes will contribute to all galaxy formation science (e.g., *JWST*, *WFIRST*, LSST, GMT, and TMT), empirical models included. For example, LSST will substantially increase coverage of dwarf galaxies and rare objects (including clusters), and GMT/TMT will provide detailed spectroscopic coverage of galaxies necessary to reduce current systematic biases of 0.2-0.3 dex in recovering many galaxy properties. Relevant to the decadal survey, we highlight two major additional needs to support empirical models.

A) **Broader, Deeper, More Complete Spectroscopic Coverage of Galaxies**

Environment (e.g., satellites vs. field galaxies) and mergers affect every galactic property, including star formation/quenching, size, kinematics, the CGM, morphology, metallicity, color, dust, and black hole mass. Empirical models will constrain the roles that environment and mergers play for all these properties, but only if observations provide environmental information. Photometric redshifts do not suffice, due to projection effects that strongly limit environmental information for individual galaxies. Currently, relatively complete (>90%) spectroscopic coverage of Milky-Way mass galaxies over a volume large enough to have many massive clusters only exists for $0<z<\sim0.2$ (SDSS and GAMA). **This short redshift baseline prevents studying how environmental effects evolve.** Studying these effects requires surveys that are:
1) **Highly complete (>90%).** Completeness often correlates both with environment (e.g., via fiber collisions) and the galaxy properties in question. At <90% completeness, selection criteria are *very* difficult to model, limiting data usability.
2) **Broad: i.e., containing a statistical sample for all environments, from voids to clusters**, as well as containing many merging (or near-merging) galaxies; otherwise, it is not possible to tell if effects exist for some environments and not others. This does not need to mean the same thing as "large area" provided that all environments/merger types are well-sampled and the footprint selection function is well-understood.
3) **Deep: specifically, such that the survey probes Milky Way-mass (or lower) galaxies to $z \geq 1$,** i.e., over a range of cosmic time such that significant galaxy evolution occurs.

The ATLAS Probe (180M galaxies to $z\sim7$ over 2000 deg$^2$; Wang et al. 2019), Maunakea Spectroscopic Explorer (800k galaxies to $z\sim3$ over 80 deg$^2$; Szeto et al. 2018) and Spectel (~8M galaxies to $z\sim4$ over 200 deg$^2$; Ellis et al. 2017) are examples that satisfy the above goals. Spectral information beyond redshifts is helpful for all galaxy properties, and it is critical for several, including the CGM, metallicities, kinematics, and AGN.  Yet, most proposed projects for the next decade cannot achieve much beyond redshifts, and so these topics may be best addressed by targeted follow-up studies with GMT/TMT.

**B) Characterization of Systematic Biases Affecting Observational Data Sets**

Observations are the product of increasingly complex software pipelines (e.g., flatfielding & calibration, PSF and profile modeling, photometric/spectroscopic redshift extraction, and stellar mass modeling).  Different pipelines have different implicit assumptions, and so it is often difficult to combine results from different projects.  Nonetheless, empirical models' constraining power increases dramatically when different observations can be combined self-consistently.  Proposed programs that satisfy the following criteria are therefore most useful:
1) **Overlap in sky area and redshift with other past and proposed surveys**, which is large enough to estimate systematic biases between methods/surveys across shared regions of galaxy parameter space.
2) **Multiple software pipelines used for all science analyses when different algorithms are available,** to allow estimating systematic pipeline-related uncertainties.
3) **Publicly accessible data for all science analyses in long-term archives.**  This allows direct comparison of the same objects and re-running new statistical analyses on archived data. It also ensures that data products can last beyond the teams that create them.
4) **Publicly available data analysis software.**  This allows teams to investigate discrepancies between results to find errors and to re-analyze results, even after original teams disband or individuals leave the field.
5) **Covariance matrices.**  Last, but not least, many observables (especially correlation functions and measures of environment) *cannot* be compared to theoretical models with any accuracy without covariance matrices.

**These criteria are crucial to ensure both <u>usability</u> and <u>legacy value</u> of datasets funded via public investment.**  As a result, programs that do not yet satisfy these criteria should be strongly encouraged to do so.

**SUMMARY**

Empirical models of galaxy formation have demonstrated both significant successes and significant promise for the next decade.  Two key abilities (revealing galaxy changes over time, and self-consistently combining multiple data sets to break degeneracies) are responsible for this promise, which is directly applicable to the most important outstanding questions in galaxy formation.  At the same time, every large-scale survey relies on mock catalogs generated from empirical models to inform survey design and validation.  To ensure continued advancement of this field, we strongly recommend investment in broad, deep, and complete galaxy redshift surveys.  We also recommend that funded programs adopt a set of practices to ensure usability and legacy value of their data—not only for empirical modeling, but for the astronomy community as a whole.  Taking these steps will continue enabling dramatic progress in both empirical modeling and galaxy evolution for the next decade.